\documentclass[aip,pof,reprint,amsmath,amssymb]{revtex4-1}
\usepackage{graphicx}
\usepackage{bm}

\begin{document}

\title{Convergent flow in a two--layer system and mountain building}

\author{Carlos Alberto Perazzo}
\email{perazzo@favaloro.edu.ar}
\thanks{Researcher of CONICET}
\affiliation{Dto. de F\'{\i}sica y Qu\'{\i}mica, Universidad Favaloro, Sol\'{\i}s 453, 1078, Buenos
Aires, Argentina.}

\author{Julio Gratton}
\email{jgratton@tinfip.lfp.uba.ar}
\thanks{Researcher of CONICET}
\affiliation{INFIP--CONICET, Dto. de F\'{\i}sica, Facultad de
Ciencias Exactas y Naturales, Universidad de Buenos Aires, Ciudad
Universitaria, Pab. I, 1428, Buenos Aires, Argentina.}

\date{\today}

\begin{abstract}
With the purpose of modelling the process of mountain building, we
investigate the evolution of the ridge produced by the convergent
motion of a system consisting of two layers of liquids that differ
in density and viscosity to simulate the crust and the upper
mantle that form a lithospheric plate. We assume that the motion
is driven by basal traction. Assuming isostasy, we derive a
nonlinear differential equation for the evolution of the thickness
of the crust. We solve this equation numerically to obtain the
profile of the range. We find an approximate self--similar
solution that describes reasonably well the process and predicts
simple scaling laws for the height and width of the range as well
as the shape of the transversal profile. We compare the
theoretical results with the profiles of real mountain belts and
find an excellent agreement.
\end{abstract}

\maketitle

\section{Introduction} \label{introduccion}
\begin{figure}
\includegraphics{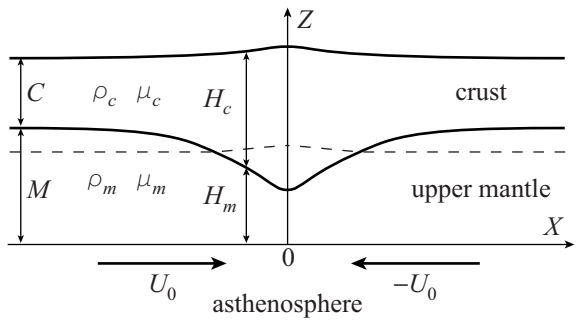}
\caption{Geometry of the two layer model employed to describe the
formation of a ridge. The line separating the crust and the upper
mantle is the Moho. The dashed line represents an isobar.}
\label{F1}
\end{figure}

Mountain ranges are one of the most striking features of the Earth
and their origin and evolution have been investigated for a long
time. It is known that the lithosphere (the outer solid layer of
the Earth) is a two layer structure in which the crust rests on
the denser upper mantle, being separated by the Mohorovi$\rm
\check{c}$i$\rm \acute{c}$ discontinuity (called Moho). The
lithosphere is divided into several approximately rigid plates
that rest on the hotter and more fluid asthenosphere. The relative
motion of these plates is the cause of mountain building, because
of the shortening and consequent thickening of the crust that
occurs when two continental plates collide (see Fig. \ref{F1} for
a sketch) or when an oceanic plate is subducted beneath a
continent. On the timescale of the orogenic processes the
lithosphere is in local hydrostatic equilibrium (a condition
called isostasy) that implies that the visible regional topography
is accompanied by a corresponding anti-topography (called root) of
the Moho.

Clearly mountain builiding is an important problem that involves
many disciplines and interests a broad range of scientists. To
attempt a realistic and detailed theoretical description of
mountain building is an exceedingly complex task (see for example
the recent review by Avouac \cite{avouac07} where the field data
are discussed) because across the lithosphere there are large
variations of the temperature, density and rheological parameters
as well as other properties (many of which, to compound the issue,
are poorly known). To this should be added the complications due
to the geometry and the time dependence of the motion of the
plates. Since the pioneering work of England and McKenzie
\cite{england82,england83} several models called collectively
`thin sheet models' that treat the lithosphere as a thin viscous
layer or layers have been developed to take into account in a
simplified way some of the above mentioned features (a
classification can be found in Refs. \onlinecite{medvedev99a,medvedev99b}).
These models have been used to describe mountain building, mainly
by means of extensive and detailed numerical simulations that deal with specific ranges.

The basic phenomena that govern the large scale evolution of
mountain belts are the spreading flow at the depth of the roots
together with isostasy and crustal shortening. The profile of the
ridge is determined by the dynamic balance between buoyancy and
viscous forces. Based on these ideas Gratton \cite{gratton89} used
dimensional arguments to derive scaling laws for the evolution of
the height and the width of a mountain belt and argued that the
evolution of the profile of a range is self--similar, even if he
could not compute the exact shape. To this purpose he estimated the viscous forces assuming that the vertical gradient of the horizontal velocity takes place near the root. However, we shall show later that this assumption is not correct, since the whole lithospheric mantle is involved in the flow. As a consequence the scaling laws of Ref.~\onlinecite{gratton89} can not describe the evolution of mountain ranges.

More recently we investigated a related problem, namely the formation of a ridge by the convergent flow of a single liquid layer over a solid moving substrate \cite{perazzo08,perazzo09,gratton09}, and found that for small
time $T$ there is a self--similar regime in which the height and
the width of the range scale as $T^{1/2}$ regardless of the
asymmetry of the flow and the rheology of the liquid. For large
time, however, a different self--similar regime is achieved in
which the height and the width follow the scaling laws obtained in Ref.
\onlinecite{gratton89}. Other researchers have also investigated
independently this problem theoretically and with a laboratory
model \cite{buck94} as well as numerically
\cite{willett99,medvedev02}.

Following our previous works we here reduce the problem to its
basic essentials, taking into account the two--layer structure of the
lithosphere but disregarding rheological and geometrical details.
For simplicity we assume a Newtonian rheology for the crust and for the lithospheric mantle, and that the problem
depends on a single horizontal cartesian coordinate. We also ignore erosion.
In this way we find approximate analytic solutions,
scaling laws and the asymptotic behavior of the process, thus
achieving a deeper physical understanding of the process.

This paper is organized as follows. In section \ref{two_layer} we
describe the assumptions and we derive the governing equations. In
section \ref{self_similar} we derive the self--similar regime
developed in the process. In section \ref{comparison} we compare
the self--similar theoretical profile with the topography of
several mountain ranges. Finally in section
\ref{discussion_and_conclusions} we discuss our work, whose main
conclusions are: (1) the simple two--layer model describes quite
well the evolution of many mountain belts, (2) their profiles have
a universal shape, and (3) to a good approximation the evolution
is self--similar, with the height and width increasing as
$T^{1/2}$.

\section{The two--layer model} \label{two_layer}
Our aim is to describe the essentials of the mountain building
process, using a model as simple as possible, in order to clarify
the basic physics involved. To this purpose we consider a two
layer liquid film as shown in Fig. \ref{F1} and we assume for
simplicity plane symmetry. The upper layer (the crust) has
viscosity $\mu_{c}$, density $\rho_{c}$ and thickness
$H_{c}(X,T)$. The lower one (the upper mantle) has viscosity
$\mu_{m}$, density $\rho_{m}$ and thickness $H_{m}(X,T)$.
Typically for a continental plate $\rho_{c} \approx 2.7$
g/cm$^{3}$, $\rho_{m} \approx 3.2$ g/cm$^{3}$, and $\mu_{c} \gg
\mu_{m}$.

Initially, both layers are uniform and $H_{c}(X,0)=C$ and
$H_{m}(X,0)=M$. To model the basal traction that is believed to
drive the plate motion we assume that at $T=0$ the bottom of the
lithosphere ($Z=0$) starts moving with a prescribed velocity
$U_{b}(X)$. We next assume isostasy,
which means that for $0 \leq Z \leq H_{m}$ (see the dashed line in
Fig. \ref{F1}) the pressure does not depend on $X$. Notice that
this implies that as the thickness of the crust increases, part of
the mass of the lithospheric mantle crosses the boundary between
the lithosphere and the asthenosphere. As a consequence the mass
of the lithospheric mantle is not conserved.

To derive the governing equations we assume a slow
viscosity--dominated flow and employ a slight generalization of
the well--known lubrication approximation (see for example Refs.
\onlinecite{buckmaster77,huppert82,oron97}) to take into account the
motion of the bottom of the lithosphere. We neglect inertia and
assume that the slope of the free surface is gentle, so that the
horizontal components of the velocities of the fluids are much
larger than the vertical ones and that their vertical gradients
are much larger than the horizontal gradients. In this way the
Stokes equation takes the form
\begin{equation}
\frac{\partial P}{\partial X} =\mu_{m}
\frac{\partial^{2}U}{\partial Z^{2}}, \quad \frac{\partial
P}{\partial Z}= \rho_{m} g, \label{stokesm}
\end{equation}
for $0 \leq Z \leq H_{m}$, and
\begin{equation}
\frac{\partial P}{\partial X} =\mu_{c}
\frac{\partial^{2}U}{\partial Z^{2}}, \quad \frac{\partial
P}{\partial Z}= \rho_{c} g, \label{stokesc}
\end{equation}
for $H_{m} \leq Z \leq H_{m} + H_{c}$. In these equations $P$ is
the pressure, $U(X,Z,T)$ is the horizontal velocity and $g$ is the
gravity. The second equations in (\ref{stokesm}) and
(\ref{stokesc}) mean that the pressure is hydrostatic; integrating
them and using the isostasy condition ($\partial P/\partial X=0$
for $0 \leq Z \leq H_{m}$) we find $\rho_{m}
\partial H_{m} / \partial X=-\rho_{c} \partial H_{c}/\partial X$.
Integrating this equation and using the initial condition we obtain
\begin{equation}
H_{m} = M + \frac{\rho_{c}}{\rho_{m}}(C-H_{c}). \label{isostasy}
\end{equation}
This allows elimination of $H_{m}$ thus yielding an equation for the
single dependent variable $H_{c}$.

To derive the velocity profile we assume that
$U(Z=0)=U_{b}$, that the velocity and the shear stress are
continuous at $Z=H_{m}$, and that the shear stress vanishes at
$Z=H_{m}+H_{c}$. Then we integrate twice the first equations in
(\ref{stokesm}) and (\ref{stokesc}) with respect to $Z$ to obtain
\begin{widetext}
\begin{equation}
U = \left\{
\begin{array}{ll}
U_b -\frac{g \rho _c }{\mu _m}H_c \left(\frac{\partial H_c}{\partial x}+\frac{\partial H_m}{\partial x}\right) Z, & 0 \leq Z \leq H_{m}\\
& \\
U_b + \frac{g \rho _c}{\mu _c}\left(\frac{\partial H_c}{\partial x}+\frac{\partial H_m}{\partial x}\right) \left[\frac{1}{2 } \left(Z-H_m\right) \left(Z-2 H_c-H_m\right)-\frac{\mu _c }{\mu _m}H_c H_m\right], & H_m \leq Z \leq H_m + H_c
           \end{array}
\right.
\label{U}
\end{equation}
\end{widetext}
Notice that the velocity profile is linear in the lithospheric mantle and parabolic in the
crust and that the average shear stress in the crust is exactly
half of that in the lithospheric mantle. This means that in most
situations the velocity drop in the crust is a small fraction of
that within the mantle. As we will show later these features of
the velocity field are crucial to determine the scaling laws for
the growth of the range.

We define the vertically averaged velocity in the
crust as
\begin{equation}
V_{c} = \frac{1}{H_c} \int_{H_m} ^{H_m + H_c} U\, dZ.
\label{Vc}
\end{equation}
We set $U_{b}(X)=U_{0}u(X)$, where $U_0$ is the maximum basal velocity so that $u$ verifies $|u|\leq 1$. Next we introduce the following dimensionless quantities
\begin{equation}
h = H_{c}/C, \quad v =V_{c}/U_{0}, \quad x =X/X_{0}, \quad t =T
U_{0}/X_{0}. \label{escalas}
\end{equation}
Here the horizontal scale $X_{0}$ is given by
\begin{equation}
X_{0}= (1-\frac{\rho_{c}}{\rho_{m}}) \frac{\rho_{c} g M
C^{2}}{\mu_{m} U_{0}}. \label{X0}
\end{equation}
Finally inserting the second of (\ref{U}) in (\ref{Vc}) and using (\ref{escalas}), we obtain
\begin{equation}
v = u - \left(1 + \frac{\rho_{c}C}{\rho_{m}M} \right) h
\frac{\partial h}{\partial x} - \frac{C}{M} \left(\frac{\mu_{m}}{3
\mu_{c}} - \frac{\rho_{c}}{\rho_{m}} \right) h^{2} \frac{\partial
h}{\partial x}. \label{vcrust}
\end{equation}
This equation together with the continuity equation
\begin{equation}
\frac{\partial h}{\partial t} + \frac{\partial (v h)}{\partial
x}=0, \label{continuity}
\end{equation}
govern the dimensionless thickness of the crust. The preceding
equations can be easily extended to two dimensions to deal with
more general geometries.

To describe the convergence of two plates we make the simplest
assumption: $u(x)=1$ for $x<0$ and $u(x)=-1$ for $x>0$. In this
way the thickness of the crust starts to increase in the region of
convergence. The initial condition is $h(x,0)=1$, and the boundary
conditions are $h(\pm \infty,t)=1$. At $x=0$ we impose the
continuity of $h$ and $v$.

\begin{figure}
\includegraphics{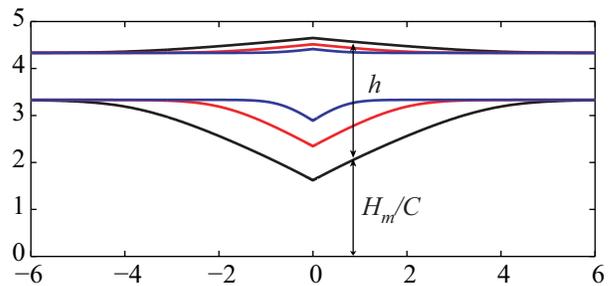}
\caption{(Color online) Numerical solutions of
(\ref{vcrust}--\ref{continuity}) with $u(x)=\pm 1$ for
$x\lessgtr0$ and $h(x,0)=1$, for $t=$0.25, 1.31 and 4.00 ($C=30$
km, $M=100$ km, $\rho_{c}=2700$ kg/m$^{3}$, $\rho_{m}=3200$
kg/m$^{3}$, $\mu_{c}/\mu_{m}=10$).} \label{F2}
\end{figure}

\begin{figure}
\includegraphics{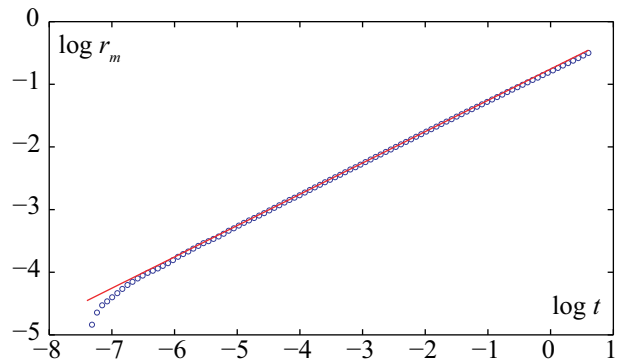}
\caption{(Color online) Evolution of the maximum relief
$r_{m}=r(0,t)$. The circles correspond to the numerical solution
with the same parameters as in Fig. \ref{F2}; the straight line is
$r_{ss}(0,t)$.} \label{F3}
\end{figure}

In general this problem must be solved
numerically. In Fig. \ref{F2} we show some solutions. All the
results shown in figures \ref{F2}, \ref{F3} and \ref{F4} were
calculated for $C=30$ km, $M=100$ km, $\rho_{c}=2700$ kg/m$^{3}$,
$\rho_{m}=3200$ kg/m$^{3}$ and $\mu_{c}/\mu_{m}=10$. These values
are representative of those found in the lithosphere, so that the
results shown can be applied in general to the mountain building
process.

\section{Self--similar regime} \label{self_similar}
We now seek the asymptotics of the problem for small $t$. We
define $r=(1-\rho_{c}/\rho_{m})(h-1)$, then $R \equiv C r$ is the
visible topography of the range. Since at the beginning of the
phenomenon $h-1 \ll 1$, the Eqs. (\ref{vcrust}) and
(\ref{continuity}) can be linearized, and with the assumption
$u(x)=\pm 1$ for $x\lessgtr0$ reduce to
\begin{equation}
\frac{\partial r}{\partial t} = \pm \frac{\partial r}{\partial x}
+ \left( 1+\alpha \right) \frac{\partial^{2} r}{\partial x^{2}},
\quad x \gtrless 0, \label{lineal2}
\end{equation}
where $\alpha=\mu_{m} C/3 \mu_{c} M$. With typical values for the
lithosphere $\alpha \approx 10^{-2}$, so that it can be neglected
and in this approximation the problem depends only on the scales
$X_{0}$, $C$ and $U_{0}$.

A solution of Eq. (\ref{lineal2}) as an infinite series similar to
that given in Ref. \onlinecite{perazzo08} exists (see the Appendix \ref{A1}). Here we shall show an approximate
self--similar solution $r_{ss}$ that for $r \ll 1$ represents the
asymptotics of the full solution. It is given by
\begin{equation}
r_{ss} \equiv \left( 1 - \frac{\rho_{c}}{\rho_{m}} \right) \frac{2
\sqrt{t}}{\sqrt{1+\alpha}} f(\psi),\label{ss}
\end{equation}
where
\begin{equation}
f(\psi)=\frac{e^{-\psi^{2}}}{\sqrt{\pi}} - \psi \,
\text{erfc}(\psi), \quad \psi \equiv \frac{x}{2 \sqrt{(1+\alpha)
t}}. \label{psi}
\end{equation}
Here erfc is the complementary error function. According to this
solution the height and the width of the ridge follow a simple
$t^{1/2}$ scaling. We define (arbitrarily) the dimensionless width
of the ridge as $w = 2 x(\psi =1)=4 \sqrt{(1+\alpha) t}$ (so the
width is the distance between the two points in which the height
is 9\% of the peak height). Then the height and the width $W=X_{0}
w $ of the ridge are given by
\begin{equation}
R = \frac{2 U_{0}}{\sqrt{\pi (1+\alpha)}} \sqrt{\left( 1 -
\frac{\rho_{c}}{\rho_{m}} \right)\frac{\mu_{m}}{\rho_{c} g M}}
T^{1/2}, \label{alto}
\end{equation}
\begin{equation}
W = \frac{4 C}{\sqrt{1+\alpha}} \sqrt{\left( 1 -
\frac{\rho_{c}}{\rho_{m}} \right) \frac{ \rho_{c} g M}{\mu_{m}}}
T^{1/2}. \label{ancho}
\end{equation}
It is interesting that $W$ depends on $C$, but not on $U_{0}$. On
the other hand $R$ depends on $U_{0}$ and is nearly independent on
$C$ (it depends on $C$ only through $\alpha$). Notice also that
the aspect ratio $\theta=W /R$ of the ridge is constant and equal
to
\begin{equation}
\theta = \frac{2 \sqrt{\pi} \rho_{c} g C M}{\mu_{m} U_{0}}.
\label{asprat}
\end{equation}
Within the uncertainties in the parameters involved, these
formulae give the correct order of magnitude of $R$ and $W$ for
real mountain ranges.

\begin{figure}
\includegraphics{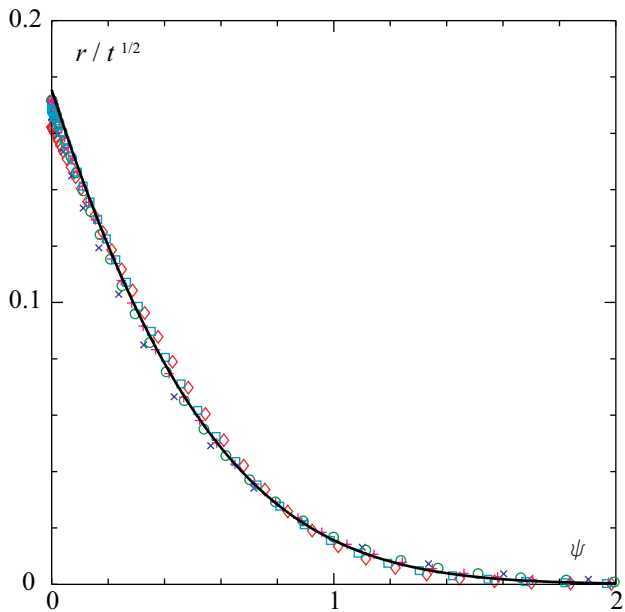}
\caption{(Color online) Scaled relief of the numerical solutions
(circles) for $t=1.4 \times 10^{-6}$ ($\times$), $5.7 \times
10^{-5}$ ($\circ$), $2.3 \times 10^{-3}$ ($+$), $9.7 \times
10^{-2}$ ($\square$) and $6.2 \times 10^{-1}$ ($\lozenge$). The
solid line is $r_{ss}/\sqrt{t}$.} \label{F4}
\end{figure}
It is interesting to compare this approximate self--similar
solution with the numerical solutions of the full nonlinear
problem (\ref{vcrust}--\ref{continuity}). In Fig. \ref{F3} we show
the numerical $r(0,t)$ and $r_{ss}(0,t)$. In Fig. \ref{F4} we
compare the numerical solutions with the solution (\ref{ss}). From
these figures it can be appreciated that the self--similar
solution (\ref{ss}--\ref{psi}) describes quite well the shape and
the evolution of the ridge, even for quite large $t$ when it might
be expected to fail (notice that the last circle of Fig. \ref{F3}
corresponds to $h(0,t=4)=3$, and that $h(0,t=0.62)=1.82$ for the
last profile in Fig. \ref{F4}). In terms of the topography this
implies that mountain ranges whose height does not exceed
approximately 5 km are well described by (\ref{ss}--\ref{psi}). We
then conclude that the self--similar solution describes reasonably
well the solution of the full nonlinear problem up to this point.
We observe that for the parameters of the numerical calculations
shown in these figures $h(0,t)=4.95$ corresponds to the root of
the ridge touching the asthenosphere, after which the relief can
not increase anymore.

\section{Comparison with real mountain ranges} \label{comparison}
It is interesting to compare the present theory with the real
profiles of mountain ranges. However at this point it is
convenient to point out that some mountain systems are not linear
so that they can not be described by the present theory. For our
comparisons we have used the digital elevation data GTOPO30
(these data are available in the website of the U.S. Geological Survey's Earth Resources Observation and Science (EROS) Center)
to obtain locally averaged profiles of 10 approximately
rectilinear segments of the Alps, Andes (2 segments), Barisan
Mountains in Sumatra, Caucasus (2 segments), New Zealand Alps,
Pyrenees and Urals (2 segments).  For each segment we have drawn
50 transversal profiles of 101 points each. All the 10 segments we
examined have the same ``pagoda roof" profile. However 4 of them
(one segment of the Andes, Caucasus and Urals, and the New Zealand
Alps) are markedly asymmetric, having one side steeper than the
other; in addition the foot of the steeper side is lower than the
other.

\begin{figure}
\includegraphics{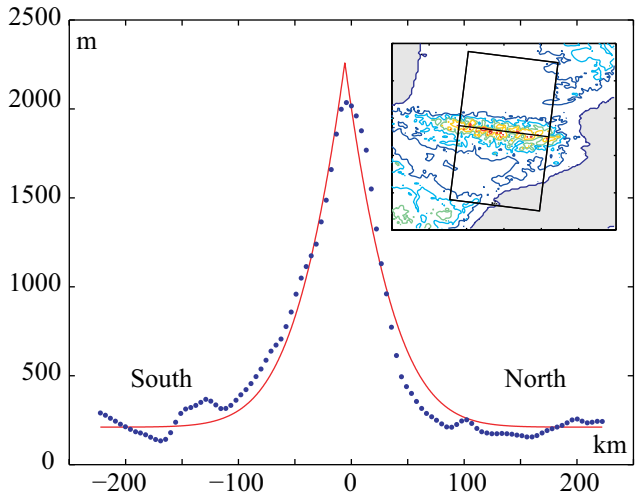}
\caption{(Color online) Comparison of the average topography
(dots) of a segment of the Pyrenees (shown in the inset) with the
theoretical profile (\ref{ss}--\ref{psi}). The full line is $a
f((X-b)/c)+d$ where $a$, $b$, $c$ and $d$ are constants lengths
determined by fitting the actual topography.} \label{F5}
\end{figure}

In Fig. \ref{F5} we show the average of the 50 profiles of a
segment of the Pyrenees along with the best fit of these data to
$a f((X-b)/c)+d$, where $f$ is given in (\ref{psi}) and $a$, $b$,
$c$ and $d$ are constant lengths.

In Fig. \ref{F6} we show the theoretical profile
(\ref{ss}--\ref{psi}) and the 6 more symmetric average profiles.
To merge these profiles in a single graph we plotted
$(R_{i}-d_{i})/a_{i}$ \emph{vs}. $(X_{i}-b_{i})/c_{i}$
($i=1,\ldots,6$). To obtain the constants $a_{i}$, $b_{i}$,
$c_{i}$, $d_{i}$ we followed the same procedure as we did for the
Pyrenees. It can be appreciated that the self--similar approximate
solution gives an excellent fit to the actual shapes.

\section{Discussion and conclusions} \label{discussion_and_conclusions}
\begin{figure}
\includegraphics{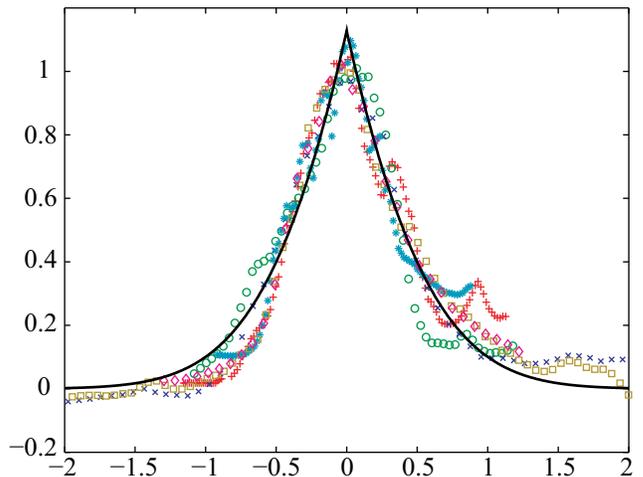}
\caption{(Color online) Comparison of the
theoretical profile from the model with the Andes ($+$), Caucasus
($\circ$), Alps ($\ast$), Urals ($\times$), Pyrenees ($\square$)
and Barisan Mountains ($\lozenge$).} \label{F6}
\end{figure}

As can be seen in figures \ref{F5} and \ref{F6} the agreement of
the profiles of actual ranges with the self--similar shape is very
good, even for a very ancient range as the Urals. However some
explanations are opportune.

The theoretical profiles are sharply peaked due to the
discontinuity of $u(x)$ at $x=0$. It is easy to solve numerically
the problem with a continuous transition of $u(x)$. We have done
it assuming that $u= \tanh(x/w_{0})$ where $2 w_{0}$ is the width of the transition. In figure \ref{F8} we compare the numerical solution for $w_{0}=0.4$ (this value was chosen for better visibility) with  the solution for the discontinuous $u$ case for the same time ($t=1.31$).  We see that a continuous transition leads to the same profile, except near the top where it is rounded. The width of this rounded region is always $\approx 2 w_{0}$, but since the width of the range increases as $t^{1/2}$ the difference between the continuous and discontinuous cases reduces with time. We conclude that the self--similar solution (\ref{ss}--\ref{psi}) describes increasingly well the profile.

The actual topographies shown in figures \ref{F5} and \ref{F6} are the result of averaging all the transversal profiles of each range. All the profiles employed to prepare these figures have a peak, but on averaging them a rounded summit is obtained. Notice that the noise present in the data due to the local topographical accidents (that occur near the surface of the crust and are not a consequence of the average lithospheric flow we are considering) introduces a horizontal scale $w_{\mathrm{noise}}$ of a few kilometers, that sets a limit to the size of the features that can be compared with the theoretical model. Then the rounded top of these figures whose sizes are of the order of $w_{\mathrm{noise}}$ do not contradict the sharp theoretical profile. In addition, this fact suggests that the transition of the basal velocity occurs on a horizontal scale shorter than $2 w_{\mathrm{noise}}$.

\begin{figure}
\includegraphics{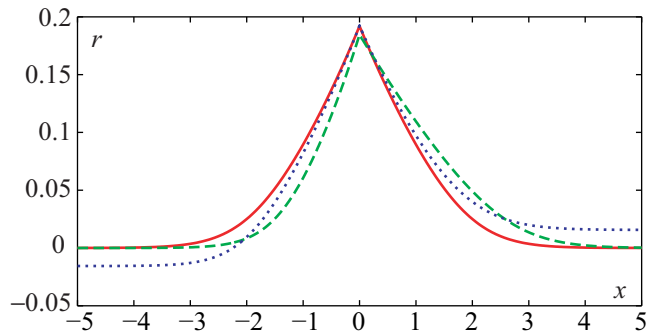}
\caption{(Color online) Comparison of the solution for symmetric case (full line) with those for nonsymmetric basal velocity (dashed line) and nonsymmetric thickness of the crust (dotted line).} \label{F7}
\end{figure}

\begin{figure}
\includegraphics{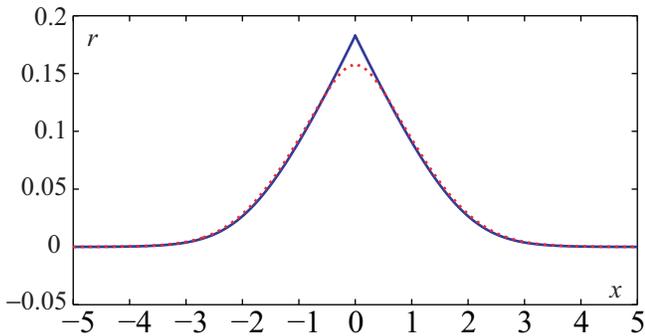}
\caption{(Color online) Comparison of a profile for a continuous basal velocity ($w_{0}=0.4$, dotted line) with the discontinuous case (full line), for $t=1.31$.} \label{F8}
\end{figure}

In our calculations we have assumed for simplicity a perfect
symmetry. However it is not difficult to extend our model to a
non--symmetric situation in which $|u|$ as well as $C$ are
different in each side of the ridge. To appreciate the effects of both kinds of asymmetries we show in figure \ref{F7} the numerical solutions for the symmetric case and those corresponding to a nonsymmetric basal velocity ($u=1.9$ for $x<0$ and $u=-0.1$ for $x>0$) and to a nonsymmetric thickness of the crust ($h(x<0,0)=0.9$ and $h(x>0,0)=1.1$), for $t=1.31$. The parameters have been chosen to ensure that in the three cases the added dimensionless mass is equal to $2t$. We can observe that regardless of the asymmetry the crest remains at $x=0$. For brevity we omit more details, that will be published elsewhere. We believe that the non--symmetric segments of the Andes, Caucasus, Urals and the New Zealand Alps can be reproduced by adequate choices of the parameters.

The present theory assumes a Newtonian rheology for the
lithosphere, although it is believed that its behavior is
non--Newtonian. In a recent article \cite{gratton09} we considered the effect of a power--law rheology in the one layer model of Ref.~\onlinecite{perazzo08}. We found that in the linear regime the maximum height and the width of the ridge increase as $t^{1/2}$ regardless of the rheological parameters. On the other hand the profile of the ridge depends on the rheology, but only weakly (see figure 4 of Ref.~\onlinecite{gratton09}). The two layer model used here can be extended to include non--Newtonian behavior but to do this exceeds the scope of the present paper. However based on the results of one layer model we expect that similar results will be obtained for the two layer model since in the linear regime both models give analogous equations.

We do not take into account in our model the effect of erosion.
Several authors have considered the role of glacial and fluvial erosion in the orogenic process,
modeling the resulting redistribution of mass at large scale as a diffusive process (see for example Refs. \onlinecite{burov07,burov09} and references therein). The inclusion in our model of this effect would modify the coefficient of the diffusion term $\partial^{2} r/\partial x^{2}$ in equation (10). This means that a
self--similar solution of the same kind as (11) and (12) would result, but with different scales. Incidentally, this could be the explanation
why our self--similar profile describes quite well all the ranges analyzed regardless of their erosion history. Notice also that this change should not modify the sharp apex of the ridge, so that a rounded summit will not result. We leave for future work a detailed investigation of the effects of erosion.

The present model can be easily generalized to include 3--D effects replacing $u$ by a two--dimensional vector $\mathbf{u} = \mathbf{u}(x,y)$ and $\partial/\partial x$ by the two--dimensional gradient operator $\nabla= (\partial/\partial x, \partial/\partial y)$. The 3--D character arises from the dependence of $\mathbf{u}$ on both cartesian coordinates. The resulting problem must then be solved numerically. The 3--D effects will be important in those parts of a range where the average curvature radius of the crest of the ridge is smaller than or of the order of its width. On the contrary, our results can be applied whenever the curvature radius is much larger than the width.

The $t^{1/2}$ scaling law can be justified with a dimensional
argument based on isostasy, conservation of the crustal mass
during the shortening and the balance between gravitational and
viscous stresses, entirely analogous to that employed in Ref. \onlinecite{gratton89}. In that paper different scaling laws were
obtained because the viscous stress was incorrectly estimated,
since it was not realized the key feature of the two layer model
dynamics, namely that the entire lithospheric mantle is involved.

Most of the papers about mountain building deal with specific ranges, chiefly the Himalaya--Tibet orogeny, that can not be described by the present model. It is interesting to compare the results of our two--layer model with those of one layer--models (see for example Medvedev \cite{medvedev02} and Perazzo and Gratton \cite{perazzo08}), and those from the two--layer model of Royden \cite{royden96}. The one layer models considers a single viscous layer on a solid horizontal substrate with convergent motion. According to Ref. \onlinecite{medvedev02,perazzo08} the height and the width of the wedge increase as $t^{\beta}$ and  $t^{1-\beta}$ respectively. In Ref. \onlinecite{medvedev02} it is found that $\beta$ decreases with time and that the evolution of the wedge can be divided into three phases. Initially, $\beta = 1$ so that the wedge grows only in height. The second phase exhibits an almost self--similar growth in which $\beta = 1/2$ so that the height and the width increase as $t^{1/2}$. For later times a last phase is achieved in which $\beta$ decreases below 0.4.  In Ref. \onlinecite{perazzo08} two self--similar regimes were found corresponding to $\beta =1/2$ for short times and to $\beta = 1/4$ for large time. In our two--layer model and for realistic values of the thickness of the lithosphere we observe only a $t^{1/2}$ self--similar regime because the root touches the asthenosphere before significant departures from this regime occur. On the other hand a $\beta = 1$ initial phase can be obtained in our two--layer model if we assume that the the basal velocity has a continuous transition whose horizontal extent is $2 w_{0}$; this phase ends around $t = \frac{4}{\pi} w_{0}^{2}$ (for brevity we omit details). Thus the one--layer and our two--layer models yield power--laws for the evolution of the height and the width which have the same exponents, notice however that the factors are quite different.

The two--layer model of Royden \cite{royden96} considers only the crust, that is divided into an upper layer with uniform viscosity and a lower layer in which the viscosity decreases exponentially with the depth. The basal traction condition is assumed to hold at the bottom of the crust. It is shown that two regimes can occur. In the first, the crustal flow is directly coupled to the underlying mantle. In the second the upper crustal to mid crustal flow is decoupled from the underlying mantle. Which one of these regimes occur depends on the viscosity just above the Moho, which in turns depends on its depth. If no significant low--viscosity zone develops, crustal deformation is coupled to the motion of the underlying mantle, and a triangular mountain range develops. If a low--viscosity zone is initially absent but develops during crustal thickening a steep--sided flat--topped plateau ultimately forms. If a low viscosity zone is present in the lower crust prior to convergence, a wide orogen with low topographic relief develops. In the last two cases crustal flow is decoupled from the mantle except at the edges of the flat region. The triangular profiles that are obtained in the coupled mode look quite similar to those obtained here. Notice however that the simplicity of our model has allowed to obtain analytic formulae for the shape of the range and its scaling laws, not previously known. Furthermore, according to our two--layer model the flow within the crust should decouple from basal traction when the root touches the asthenosphere, being driven only by gravity, possibly yielding a flat--topped profile similar to those discussed in Ref. \onlinecite{royden96}. We have not yet investigated this regime.

We conclude that the simple two layer model describes quite well
the evolution of many mountain belts. Although the lithosphere is
described by many parameters, to a good approximation the orogenic
process involves only $U_{0}$, $C$ and the combination $X_{0}$
(Eq. (\ref{X0})). Furthermore as long as $\mu_{m} \ll \mu_{c}$ the
viscosity of the crust is not relevant, since most of the vertical
gradient of the velocity occurs in the lithospheric mantle. The
evolution of mountain belts is to good approximation self--similar
and in the symmetric case the profile is given by
(\ref{ss}--\ref{psi}).

\begin{acknowledgments}
We acknowledge grant PICTO FONCYT/UF 21360 BID OC/AR 1728 from
FONCYT and Universidad Favaloro.
\end{acknowledgments}
\appendix*
\section{Linearized series solution}
\label{A1}
Introducing the scaled variables
\[
\tilde{t} = t/(1+\alpha), \quad \tilde{x} = x/(1+\alpha), \quad \tilde{r} = r/(1+\alpha)
\]
in equation (\ref{lineal2}) and following the procedure described in the Appendix of Ref.~\onlinecite{perazzo08}, we obtain the solution for $x>0$ as
\begin{equation}
\tilde{r} = \frac{e^{-s^2}}{\sqrt{\pi}}\left[2\sqrt{\tilde{t}} \, H_{-2}(s)+
\sum _{j=1}^{\infty} \left(2 \sqrt{\tilde{t}}\,\right)^j H_{-1-j}(s) \right],
\label{ape1}
\end{equation}
where $s=(\tilde{t} + \tilde{x})/2 \sqrt{\tilde{t}}$ and $H_{q}(s)$ denotes the
Hermite function of order $q$. To obtain the solution for $x<0$ one must change $x$ for $-x$ in (\ref{ape1}).

\begin{thebibliography}{10}%
\makeatletter
\providecommand \@ifxundefined [1]{%
 \ifx #1\undefined \expandafter \@firstoftwo
 \else \expandafter \@secondoftwo
\fi
}%
\providecommand \@ifnum [1]{%
 \ifnum #1\expandafter \@firstoftwo
 \else \expandafter \@secondoftwo
\fi
}%
\providecommand \enquote [1]{``#1''}%
\providecommand \bibnamefont  [1]{#1}%
\providecommand \bibfnamefont [1]{#1}%
\providecommand \citenamefont [1]{#1}%
\providecommand\href[0]{\@sanitize\@href}%
\providecommand\@href[1]{\endgroup\@@startlink{#1}\endgroup\@@href}%
\providecommand\@@href[1]{#1\@@endlink}%
\providecommand \@sanitize [0]{\begingroup\catcode`\&12\catcode`\#12\relax}%
\@ifxundefined \pdfoutput {\@firstoftwo}{%
 \@ifnum{\z@=\pdfoutput}{\@firstoftwo}{\@secondoftwo}%
}{%
 \providecommand\@@startlink[1]{\leavevmode}%
 \providecommand\@@endlink[0]{}%
}{%
 \providecommand\@@startlink[1]{%
  \leavevmode
  \pdfstartlink
   attr{/Border[0 0 1 ]/H/I/C[0 1 1]}%
   user{/Subtype/Link/A<</Type/Action/S/URI/URI(#1)>>}%
  \relax
 }%
 \providecommand\@@endlink[0]{\pdfendlink}%
}%
\providecommand \url  [0]{\begingroup\@sanitize \@url }%
\providecommand \@url [1]{\endgroup\@href {#1}{\urlprefix}}%
\providecommand \urlprefix [0]{URL }%
\providecommand \Eprint[0]{\href }%
\@ifxundefined \urlstyle {%
  \providecommand \doi [1]{doi:\discretionary{}{}{}#1}%
}{%
  \providecommand \doi [0]{doi:\discretionary{}{}{}\begingroup
  \urlstyle{rm}\Url }%
}%
\providecommand \doibase [0]{http://dx.doi.org/}%
\providecommand \Doi[1]{\href{\doibase#1}}%
\providecommand \selectlanguage [0]{\@gobble}%
\providecommand \bibinfo [0]{\@secondoftwo}%
\providecommand \bibfield [0]{\@secondoftwo}%
\providecommand \translation [1]{[#1]}%
\providecommand \BibitemOpen[0]{}%
\providecommand \bibitemStop [0]{}%
\providecommand \bibitemNoStop [0]{.\EOS\space}%
\providecommand \EOS [0]{\spacefactor3000\relax}%
\providecommand \BibitemShut [1]{\csname bibitem#1\endcsname}%
\bibitem{avouac07}%
  \BibitemOpen
  \bibfield{author}{%
  \bibinfo {author} {\bibfnamefont{J.~P.}\ \bibnamefont{Avouac}},\ }%
  \enquote{\bibinfo {title} {Dynamic processes in extensional and compressional
  settings -- mountain building: from earthquakes to geological deformation},}\
  in\ \emph{\bibinfo {booktitle} {Crust and lithosphere dynamics, Treatise on
  Geophysics Vol. 6}},\ \bibinfo {editor} {edited by\ \bibinfo {editor}
  {\bibfnamefont{G.}~\bibnamefont{Schubert}}}\ (\bibinfo {publisher}
  {Elsevier},\ \bibinfo {address} {Cambridge},\ \bibinfo {year} {2007})\ pp.\ \bibinfo {pages}
  {377--439}\BibitemShut{NoStop}%
\bibitem{england82}%
  \BibitemOpen
  \bibfield{author}{%
  \bibinfo {author} {\bibfnamefont{P.}~\bibnamefont{{England}}}\ and\ \bibinfo
  {author} {\bibfnamefont{D.}~\bibnamefont{{McKenzie}}},\ }%
  \bibfield{title}{%
  \enquote{\bibinfo {title} {{A thin viscous sheet model for continental
  deformation}},}\ }%
  \bibfield{journal}{%
  \Doi{10.1111/j.1365-246X.1982.tb04969.x}{\bibinfo {journal} {Geophys. J.
  Internat.}}\ }%
  \textbf{\bibinfo {volume} {70}},\ \bibinfo {pages} {295--321} (\bibinfo
  {year} {1982})\BibitemShut{NoStop}%
\bibitem{england83}%
  \BibitemOpen
  \bibfield{author}{%
  \bibinfo {author} {\bibfnamefont{P.}~\bibnamefont{{England}}}\ and\ \bibinfo
  {author} {\bibfnamefont{D.}~\bibnamefont{{McKenzie}}},\ }%
  \bibfield{title}{%
  \enquote{\bibinfo {title} {{Correction to: a thin viscous sheet model for
  continental deformation}},}\ }%
  \bibfield{journal}{%
  \Doi{10.1111/j.1365-246X.1983.tb03328.x}{\bibinfo {journal} {Geophys. J.
  Internat.}}\ }%
  \textbf{\bibinfo {volume} {73}},\ \bibinfo {pages} {523--532} (\bibinfo
  {year} {1983})\BibitemShut{NoStop}%
\bibitem{medvedev99a}%
  \BibitemOpen
  \bibfield{author}{%
  \bibinfo {author} {\bibfnamefont{S.~E.}\ \bibnamefont{{Medvedev}}}\ and\
  \bibinfo {author} {\bibfnamefont{Y.~Y.}\ \bibnamefont{{Podladchikov}}},\ }%
  \bibfield{title}{%
  \enquote{\bibinfo {title} {{New extended thin-sheet approximation for
  geodynamic applications-I. Model formulation}},}\ }%
  \bibfield{journal}{%
  \Doi{10.1046/j.1365-246x.1999.00734.x}{\bibinfo {journal} {Geophys. J.
  Internat.}}\ }%
  \textbf{\bibinfo {volume} {136}},\ \bibinfo {pages} {567--585} (\bibinfo
  {year} {1999})\BibitemShut{NoStop}%
\bibitem{medvedev99b}%
  \BibitemOpen
  \bibfield{author}{%
  \bibinfo {author} {\bibfnamefont{S.~E.}\ \bibnamefont{{Medvedev}}}\ and\
  \bibinfo {author} {\bibfnamefont{Y.~Y.}\ \bibnamefont{{Podladchikov}}},\ }%
  \bibfield{title}{%
  \enquote{\bibinfo {title} {{New extended thin-sheet approximation for
  geodynamic applications-II. Two-dimensional examples}},}\ }%
  \bibfield{journal}{%
  \Doi{10.1046/j.1365-246x.1999.00735.x}{\bibinfo {journal} {Geophys. J.
  Internat.}}\ }%
  \textbf{\bibinfo {volume} {136}},\ \bibinfo {pages} {586--608} (\bibinfo
  {year} {1999})\BibitemShut{NoStop}%
\bibitem{gratton89}%
  \BibitemOpen
  \bibfield{author}{%
  \bibinfo {author} {\bibfnamefont{J.}~\bibnamefont{{Gratton}}},\ }%
  \bibfield{title}{%
  \enquote{\bibinfo {title} {{Crustal shortening, root spreading, isostasy, and
  the growth of orgenic belts: A dimensional analysis}},}\ }%
  \bibfield{journal}{%
  \Doi{10.1029/JB094iB11p15627}{\bibinfo {journal} {J. Geophys. Res.}}\ }%
  \textbf{\bibinfo {volume} {94}},\ \bibinfo {pages} {15627--15634} (\bibinfo
  {year} {1989})\BibitemShut{NoStop}%
\bibitem{perazzo08}%
  \BibitemOpen
  \bibfield{author}{%
  \bibinfo {author} {\bibfnamefont{C.~A.}\ \bibnamefont{{Perazzo}}}\ and\
  \bibinfo {author} {\bibfnamefont{J.}~\bibnamefont{{Gratton}}},\ }%
  \bibfield{title}{%
  \enquote{\bibinfo {title} {{Asymptotic regimes of ridge and rift formation in
  a thin viscous sheet model}},}\ }%
  \bibfield{journal}{%
  \Doi{10.1063/1.2908356}{\bibinfo {journal} {Phys. of Fluids}}\ }%
  \textbf{\bibinfo {volume} {20}},\ \bibinfo {pages} {043103} (\bibinfo {year}
  {2008})\BibitemShut{NoStop}%
\bibitem{perazzo09}%
  \BibitemOpen
  \bibfield{author}{%
  \bibinfo {author} {\bibfnamefont{C.~A.}\ \bibnamefont{{Perazzo}}}\ and\
  \bibinfo {author} {\bibfnamefont{J.}~\bibnamefont{{Gratton}}},\ }%
  \bibfield{title}{%
  \enquote{\bibinfo {title} {{Self--similar asymptotics in non--symmetrical
  convergent viscous gravity currents}},}\ }%
  \bibfield{journal}{%
  \Doi{10.1088/1742-6596/166/1/012012}{\bibinfo {journal} {J. Phys.: Conf.
  Ser.}}\ }%
  \textbf{\bibinfo {volume} {166}},\ \bibinfo {pages} {012012--+} (\bibinfo
  {year} {2009})\BibitemShut{NoStop}%
\bibitem{gratton09}%
  \BibitemOpen
  \bibfield{author}{%
  \bibinfo {author} {\bibfnamefont{J.}~\bibnamefont{{Gratton}}}\ and\ \bibinfo
  {author} {\bibfnamefont{C.~A.}\ \bibnamefont{{Perazzo}}},\ }%
  \bibfield{title}{%
  \enquote{\bibinfo {title} {{Self--similar asymptotics in convergent viscous
  gravity currents of non--Newtonian liquids}},}\ }%
  \bibfield{journal}{%
  \Doi{10.1088/1742-6596/166/1/012011}{\bibinfo {journal} {J. Phys.: Conf.
  Ser.}}\ }%
  \textbf{\bibinfo {volume} {166}},\ \bibinfo {pages} {012011--+} (\bibinfo
  {year} {2009})\BibitemShut{NoStop}%
\bibitem{buck94}%
  \BibitemOpen
  \bibfield{author}{%
  \bibinfo {author} {\bibfnamefont{W.~R.}\ \bibnamefont{{Buck}}}\ and\ \bibinfo
  {author} {\bibfnamefont{D.}~\bibnamefont{{Sokoutis}}},\ }%
  \bibfield{title}{%
  \enquote{\bibinfo {title} {{Analogue Model of Gravitational Collapse and
  Surface Extension during Continental Convergence}},}\ }%
  \bibfield{journal}{%
  \Doi{10.1038/369737a0}{\bibinfo {journal} {Nature}}\ }%
  \textbf{\bibinfo {volume} {369}},\ \bibinfo {pages} {737--740} (\bibinfo
  {year} {1994})\BibitemShut{NoStop}%
\bibitem{willett99}%
  \BibitemOpen
  \bibfield{author}{%
  \bibinfo {author} {\bibfnamefont{S.}~\bibnamefont{{Willett}}},\ }%
  \bibfield{title}{%
  \enquote{\bibinfo {title} {{Rheological dependence of extension in wedge
  models of convergent orogens}},}\ }%
  \bibfield{journal}{%
  \Doi{10.1016/S0040-1951(99)00034-7}{\bibinfo {journal} {Tectonophysics}}\ }%
  \textbf{\bibinfo {volume} {305}},\ \bibinfo {pages} {419--435} (\bibinfo
  {year} {1999})\BibitemShut{NoStop}%
\bibitem{medvedev02}%
  \BibitemOpen
  \bibfield{author}{%
  \bibinfo {author} {\bibfnamefont{S.}~\bibnamefont{{Medvedev}}},\ }%
  \bibfield{title}{%
  \enquote{\bibinfo {title} {{Mechanics of viscous wedges: Modeling by
  analytical and numerical approaches}},}\ }%
  \bibfield{journal}{%
  \Doi{10.1029/2001JB000145}{\bibinfo {journal} {J. Geophys. Res.}}\ }%
  \textbf{\bibinfo {volume} {107}},\ \bibinfo {pages} {2123--+} (\bibinfo
  {year} {2002})\BibitemShut{NoStop}%
\bibitem{buckmaster77}%
  \BibitemOpen
  \bibfield{author}{%
  \bibinfo {author} {\bibfnamefont{J.~D.}\ \bibnamefont{{Buckmaster}}},\ }%
  \bibfield{title}{%
  \enquote{\bibinfo {title} {{Viscous sheets advancing over dry beds}},}\ }%
  \bibfield{journal}{%
  \Doi{10.1017/S0022112077002328}{\bibinfo {journal} {J. Fluid Mech.}}\ }%
  \textbf{\bibinfo {volume} {81}},\ \bibinfo {pages} {735--756} (\bibinfo
  {year} {1977})\BibitemShut{NoStop}%
\bibitem{huppert82}%
  \BibitemOpen
  \bibfield{author}{%
  \bibinfo {author} {\bibfnamefont{H.~E.}\ \bibnamefont{{Huppert}}},\ }%
  \bibfield{title}{%
  \enquote{\bibinfo {title} {{The propagation of two-dimensional and
  axisymmetric viscous gravity currents over a rigid horizontal surface}},}\ }%
  \bibfield{journal}{%
  \Doi{10.1017/S0022112082001797}{\bibinfo {journal} {J. Fluid Mech.}}\ }%
  \textbf{\bibinfo {volume} {121}},\ \bibinfo {pages} {43--58} (\bibinfo {year}
  {1982})\BibitemShut{NoStop}%
\bibitem{oron97}%
  \BibitemOpen
  \bibfield{author}{%
  \bibinfo {author} {\bibfnamefont{Alexander}\ \bibnamefont{Oron}}, \bibinfo
  {author} {\bibfnamefont{Stephen~H.}\ \bibnamefont{Davis}},\ and\ \bibinfo
  {author} {\bibfnamefont{S.~George}\ \bibnamefont{Bankoff}},\ }%
  \bibfield{title}{%
  \enquote{\bibinfo {title} {Long-scale evolution of thin liquid films},}\ }%
  \bibfield{journal}{%
  \Doi{10.1103/RevModPhys.69.931}{\bibinfo {journal} {Rev. Mod. Phys.}}\ }%
  \textbf{\bibinfo {volume} {69}},\ \bibinfo {pages} {931--980} (\bibinfo
  {month} {Jul}\ \bibinfo {year} {1997})\BibitemShut{NoStop}%
\bibitem{burov07}%
  \BibitemOpen
  \bibfield{author}{%
  \bibinfo {author} {\bibfnamefont{E.}~\bibnamefont{Burov}}\ and\ \bibinfo
  {author} {\bibfnamefont{G.}~\bibnamefont{Toussaint}},\ }%
  \bibfield{title}{%
  \enquote{\bibinfo {title} {Surface processes and tectonics: Forcing of
  continental subduction and deep processes},}\ }%
  \bibfield{journal}{%
  \Doi{DOI: 10.1016/j.gloplacha.2007.02.009}{\bibinfo {journal} {Global and
  Planetary Change}}\ }%
  \textbf{\bibinfo {volume} {58}},\ \bibinfo {pages} {141 -- 164} (\bibinfo
  {year} {2007})\BibitemShut{NoStop}%
\bibitem{burov09}%
  \BibitemOpen
  \bibfield{author}{%
  \bibinfo {author} {\bibfnamefont{E.}~\bibnamefont{Burov}},\ }%
  \enquote{\bibinfo {title} {Thermo--mechanical models for coupled
  lithosphere--surface processes: Applications to continental convergence and
  mountain building processes},}\ in\ \emph{\bibinfo {booktitle} {New Frontiers
  in Integrated Solid Earth Sciences}}\ (\bibinfo {publisher} {Springer
  Netherlands},\ \bibinfo {address} {Cambridge},\ \bibinfo {year} {2009})\ pp.\
  \bibinfo {pages} {103--143}\BibitemShut{NoStop}%
\bibitem{royden96}%
  \BibitemOpen
  \bibfield{author}{%
  \bibinfo {author} {\bibfnamefont{L.}~\bibnamefont{{Royden}}},\ }%
  \bibfield{title}{%
  \enquote{\bibinfo {title} {{Coupling and decoupling of crust and mantle in
  convergent orogens: Implications for strain partitioning in the crust}},}\ }%
  \bibfield{journal}{%
  \Doi{10.1029/96JB00951}{\bibinfo {journal} {J. Geophys. Res. (Solid Earth)}}\
  }%
  \textbf{\bibinfo {volume} {101}},\ \bibinfo {pages} {17679--17706} (\bibinfo
  {year} {1996})\BibitemShut{NoStop}%
\end{thebibliography}
\end{document}